\documentclass[conference,10pt]{IEEEtran}

\usepackage{amssymb,amsmath,amsfonts,bm,epsfig,graphicx,theorem,latexsym}
\usepackage{rotating,setspace,latexsym,epsf,color,subfigure}
\usepackage{cite,authblk}

\newtheorem{Lemma}{Lemma}

\def \n2{{N_0 \over 2}}

\def \h5{\hspace{0.5in}}

\begin{document}
\IEEEoverridecommandlockouts
\pagestyle{empty}

\title{On Age and Value of Information in Status Update Systems \vspace{-0.05in}}

\author{Peng Zou \qquad Omur Ozel \qquad Suresh Subramaniam \\
\normalsize Department of Electrical and Computer Engineering \\
\normalsize George Washington University, Washington, DC 20052 USA \\
\normalsize {\it pzou94, ozel, suresh@gwu.edu} \vspace{-0.1in}}

\maketitle 

\begin{abstract}
Motivated by the inherent value of packets arising in many cyber-physical applications (e.g., due to precision of the information content or an alarm message), we consider status update systems with update packets carrying values as well as their generation time stamps. Once generated, a status update packet has a random initial value and a deterministic deadline after which it is not useful (ultimate staleness). In our model, value of a packet decreases in time (even after reception) starting from its generation to ultimate staleness when it vanishes. The value of information (VoI) at the receiver is \textit{additive} in that the VoI is the sum of the current values of all packets held by the receiver. We investigate various queuing disciplines under potential dependence between value and service time and provide closed form expressions for average VoI at the receiver. Numerical results illustrate the average VoI for different scenarios and the contrast between average age of information (AoI) and average VoI. 
\end{abstract}\vspace{-0.05in}

\section{Introduction}

In many cyber-physical applications, the need for \textit{real-time} communication of information packets involves maintaining information freshness and is accompanied by the need to assign values for those packets. Examples of such cases include autonomous cars and general vehicular networks \cite{giordani2019investigating,higuchi2019value,giordani2019framework}, sensor networks \cite{boloni2013scheduling,turgut2013ive,bidoki2018joint}, tactical networks \cite{suri2015exploring} and other systems making decisions in \textit{real-time} \cite{kamar2013light, rosenthal2013look}. In this context, value of information (VoI) measuring how important gathered packets at a receiver are for applications that use them is a natural and relatively unexplored dimension to the information freshness of update packets. In this paper, we address this issue in queuing systems carrying status update packets. 

Status update systems with the age of information (AoI) metric measuring end-to-end freshness of packets have received recent active research interest. Pioneered by the analysis in \cite{kaul2012real,kaul2011minimizing} motivated from vehicular status update systems, the AoI metric has been found useful in various scenarios such as single server queuing systems \cite{costa2016age,najm2016age,inoue2018general}, energy harvesting systems \cite{2018information, bacinoglu2015age, yates2015lazy, wu2017optimal_ieee, bacinoglu2017scheduling, farazi2018average}, single and multi-hop networks \cite{arafa2017age, talak2017minimizing, kadota2018optimizing, yates2018age, maatouk2018age} and vehicular communication networks \cite{Alabbasi2018JointIF}. AoI metric gives exclusive meaning to the timing of packets and connects a packet's usefulness at the receiver with how long the packet spends before its reception. As such, each packet is assumed to be created with the same value starting at generation. 
Current literature on status update system abstractions is focused mostly on information freshness and does not exclusively consider real-time communication of information packets involving a (time varying) value associated with its content as well as timing with some recent attempts in \cite{rajaraman2018not,AKosta2017,maatouk2019age,bastopcu2019,ayan2019age} as exceptions. Different packets generated at different times may have different values. In such cases, AoI metric falls short of capturing all dimensions of the problem and a separate value of information (VoI) has to be introduced. 

In this paper, we abstract out the value of information (VoI) of a status update packet as a time-varying quantity with a random initial value and that decreases to zero within a deterministic deadline (identical over all packets). Packets are assumed useless after the deadline, which we term ultimate staleness. We also assume a functional dependence between the initial value of an information packet and its service time addressing relations between value and data sizes (e.g., packets carrying higher resolution information are more valuable but larger in size), the growth rate of processes to be monitored (e.g., state estimation in cyber-physical systems) or the content of packets regarding an alarming event. Different from the AoI metric, VoI is \textit{additive} since packets received at different instants are collected at the receiver, reminiscent of throughput (c.f. \cite{kadota2018optimizing} for a comparison of throughput and AoI). The VoI at the receiver is the sum of the values of all the packets currently held by the receiver. Note that the value of a packet continues to decay after it is received until ultimate staleness is reached. 

We note that the use of deadlines has been a topic of research in earlier works in the literature on AoI, motivating us to further explore it in the context of value of information updates. Reference \cite{CKam2016MILCOM} shows how packet deadlines, buffer sizes and packet replacement influence average AoI. Closed-form expressions for average AoI with deadline are derived in \cite{CKam2016ISIT,CKam2018}. Reference \cite{YInoue2018} studies AoI in a status update system with random packet deadlines and infinite buffer capacity. 


Previous work in \cite{rajaraman2018not,bastopcu2019,maatouk2019age,AKosta2017,ayan2019age} have components related to our view on value of information. For example, references \cite{rajaraman2018not,bastopcu2019} consider quality of information associated with the distortion observed at the receiving end. Similarly, \cite{maatouk2019age} relates timeliness of observations with correctness of information. \cite{AKosta2017} considers age and value of information with a notion of value taking into account the non-linear costs regarding information updates in various queuing disciplines. The work in \cite{ayan2019age} evaluates value of information in addition to age of information in uplink/downlink transmissions in network control systems. In the current paper, we propose a new notion of VoI compared to existing works where a packet's inherent properties at the time of generation determines its value in contrast to a value evaluated after processing at the receiver. Each packet's value decreases in time until a predetermined deadline. We investigate VoI in M/G/1/1, M/G/1/2 and M/G/1/$2^*$ queuing disciplines and we provide closed form expressions for average VoI to compare it with AoI. Our numerical results show average VoI figures for each case with various functional dependencies and the contrast between average AoI and average VoI.

\section{System Model}

\begin{figure}[!t]
	\centering{
		\hspace{-0.0cm} 
		\includegraphics[totalheight=0.08\textheight]{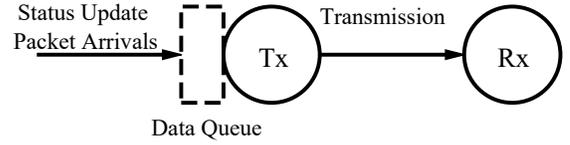}}\vspace{-0.1in}
	\caption{\sl System model with status update packets arriving to a single server	transmission queue.}
	\label{fig:1} 
	\vspace{-0.25in}
\end{figure}

We consider a point-to-point communication system with a single transmitter-receiver pair sending status updates from a single source as shown in Fig. \ref{fig:1}. The update packets arrive at the transmitter as a Poisson process with arrival rate $\lambda$ at instants $t_i$. A packet may be discarded in the queuing phase; those that are not discarded enter the server and are received by the receiver after service time $S_i$ at $t_i' = t_i + S_i$. Here, $S_i$ is independent and identically distributed with $f_S(s)$. In this paper, we cover M/GI/1/1, M/GI/1/2 and M/GI/1/$2^*$ queuing schemes. In M/GI/1/1, there is no buffer and packets arriving in busy state are discarded. In M/GI/1/2 and M/GI/1/$2^*$, there is a single data buffer with first come first serve and last come first serve disciplines, respectively. 

\subsection{Value of a Packet}

The $i$th update packet has initial value $V_{0,i}$ at the generation instant. This is a random sequence independent over different $i$. $V_{0,i}$ has the identical general distribution $f_V(v)$ with mean value $\mathbb{E}[V]$. This initial value represents the importance of a packet for an application. It could be related to the precision of a measurement, proximity of the sensor to the measured object, or it could indicate an alarm event. Each packet has a deterministic lifetime $D$ after which it reaches ultimate staleness. Hence after a fixed time period $D$ from the packet generation, the packet has no value for the receiver. 

Motivated by various applications of sensor networking and the value of information in them \cite{giordani2019investigating,higuchi2019value,giordani2019framework,boloni2013scheduling,turgut2013ive,bidoki2018joint}, in our model, we assume that the packet $i$'s value decreases from time of generation at $t_i$ until it hits deadline at $t_i+D$. The value $V_i(\tau)=f(\tau)$ for $i$th packet decreases with $\tau=t-t_i$ representing the time passed after generation at the transmitter. This value keeps on decreasing (even after a packet is received) until it becomes zero. We have $f(0)=V_{0,i}$ and $f(D)=0$. We consider different descend functions $f(.)$ for the value under three different categories: (i) concave descend, (ii) convex descend and (iii) linear descend. In the linear case, for example, since $f(0)=V_{0,i}$ and $f(D)=0$, we have the linear descend function:
\begin{align}\label{LinearD}
V_i(\tau)=f(\tau)=-\frac{V_{0,i}}{D}\tau+V_{0,i}.
\end{align}
We can also use the boundary conditions to obtain specific forms of convex and concave $f$ functions.

\subsection{Value-Dependent Service Times}

In our model, the service time of a packet depends on the initial value of the packet through a function $g$:
\begin{align}
S_i=g(V_{0,i}).
\end{align}
In this case, the distribution function of $S_i$ is $f_S(s)=f_V(g^{-1}(s))\frac{dg^{-1}(s)}{ds}$
where $g^{-1}(.)$ is the inverse function of $g(.)$, and the mean service time is
$\mathbb{E}[S]=\mathbb{E}[g(V)]$. This monotonic relation reflects the fact that a larger packet takes longer time to transmit and its reception yields more value. This relation causes a tradeoff between value and age as larger value is obtained at the receiver by paying a longer service time.

A crucial difference of VoI with respect to AoI is that the receiver \textit{collects} VoI of received packets; cf. \cite{boloni2013scheduling,turgut2013ive,bidoki2018joint} where additive nature of VoI is discussed in various wireless sensor networks. Hence, instantaneous VoI is:
\begin{align}
VoI(t) = \sum_{j=1}^{i_t} V_i(t)
\end{align}
where $i_t = \max\{i: t'_i \leq t\}$. 

\begin{figure}[!t]
	\centering{
		\hspace{-0.0cm} 
		\includegraphics[totalheight=0.2\textheight]{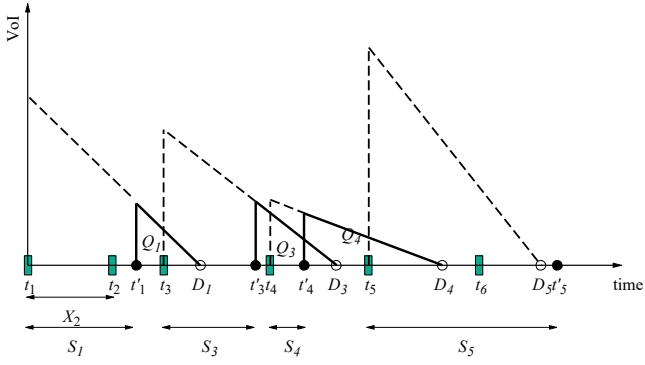}}\vspace{-0.1in}
	\caption{\sl Evolution of value for packets in M/GI/1/1 scheme.}
	\label{evol:MG11} 
	\vspace{-0.25in}
\end{figure}

In Fig. \ref{evol:MG11}, the evolution of value for specific packets generated over time is shown in M/GI/1/1 with linear descend function. Packet 1 finds the server idle and begins service at $t_1$; service ends at $t_1'$. Between $t_1$ and $t_1'$, packet 2 arrives and is discarded. The service of packet $1$ finishes at $t_1'$ before the deadline of packet 1, $D_1=t_1+D$. The value of packet 1 at $t_1'$ when received by the receiver is non-zero and it keeps descending to zero until $D_1$. Packets 3, 4 and 5 arrive to the system during idle period and are received at $t_3'$, $t_4'$, $t_5'$. Note that once packet 4 is received, packet 3 has non-zero value and the instantaneous value obtained at the receiver is the sum of them. 

We define areas $Q_i$ under the triangular regions of the curve shown in Fig. \ref{evol:MG11} and we set $Q_i=0$ for packets discarded in the queuing phase. Then the time average VoI at the receiver is:
\begin{align}\label{avoi}
\mathbb{E}[ VoI ]=\lambda\mathbb{E}[Q_i],
\end{align}
where $\lambda$ is the arrival rate for the system.

\section{Evaluating VoI for Linear Descend Function}

In this section, we devise closed form expressions for $\mathbb{E}[VoI]$ for various packet management schemes.

\subsection{Average VoI for $M/GI/1/1$}

In $M/GI/1/1$ queueing system, there is a single server and no buffer. Packets that arrive in the idle period are taken to service immediately and those arriving in busy period are dropped. In view of the renewal structure, we have the following stationary probabilities for each state:
\begin{equation}\label{MG11:pb}
p_{I}=\frac{1}{\lambda T_{cycle}},\ p_{B}=\frac{\mathbb{E}[S]}{T_{cycle}},
\end{equation}
where $T_{cycle}=\frac{1}{\lambda}+\mathbb{E}[S]$ is the expected length of one renewal cycle; $I$ and $B$ indicate the idle and busy states. Recall that if total time spent before reaching destination is larger than $D$, its value vanishes. Since a packet that is taken to service spends service time $S_i$ in the queue before reaching destination, that packet's value vanishes if $S_{i}$ is larger than $D$. Hence, we just need to consider the condition $S_i<D$ and $i$ arriving in idle state, and we get:
\begin{align}\label{EQi:MG11}
\nonumber Q_{i}&=(V_{0,i}-\frac{V_{0,i}}{D}S_{i})(D-S_{i})/2 \\
\mathbb{E}[Q_{i}]&=\frac{p_I}{2}\int_{0}^{\tilde{V}}\frac{v}{D}(D-g(v))^2f_{V}(v)dv,
\end{align}
where $\tilde{V}=g^{-1}(D)$. Then, the time average value at the receiver is $\mathbb{E}[VoI]=\lambda \mathbb{E}[Q_{i}]$.

\subsection{Average Value for $M/GI/1/2$}

In $M/GI/1/2$ queueing system, there is a single buffer. The server is in either idle or busy state. Packets that arrive in the idle period are serviced immediately and those that arrive in busy period are stored in the buffer if there is no other packet in it. They are discarded otherwise. In view of the renewal structure, we have the following stationary probabilities for each state of the server:
\begin{align}\label{prob:MG12}
p_{I}=\frac{1}{\lambda T_{cycle}},\ p_{B}=\frac{\mathbb{E}[S]}{T_{cycle}MGF_{S}(\lambda)},
\end{align}
where we use $MGF_{S}(\lambda)$ to denote the moment generating function of the service distribution evaluated at $-\lambda$:
\begin{align}
MGF_{S}(\lambda)= \mathbb{E}[e^{-\lambda S}],
\end{align} 
where $T_{cycle}=\frac{1}{\lambda}+\frac{\mathbb{E}[S]}{MGF_{S}(\lambda)}$ is the expected length of one renewal cycle. Next, we evaluate $\mathbb{E}[Q_{i}| (s)]$ for $s \in \mathcal{S}_{M/GI/1/2}=\{I,B\}$ and conditioning is on the server state observed by packet $i$. Due to PASTA property, $\mbox{Pr}[P_{i}=(s)]=p_{s}$ where $p_s$, $s \in \mathcal{S}_{M/GI/1/2}$ are as in (\ref{prob:MG12}).
\subsubsection{$\mathbb{E}[Q_{i}|I]$} As a packet arriving in idle state is served immediately, we have:
\begin{align}\label{EQ,I:MG12}
\nonumber Q_{i}=\left\{\begin{matrix}
\frac{V_{0,i}}{2D}(D-S_{i})^2 \ \ &(S_{i}<D) \\ 
0 \ \ &(S_{i}>D) 
\end{matrix}\right.\\
 \mathbb{E}[Q_{i}|I]=\frac{1}{2}\int_{0}^{\tilde{V}}\frac{v}{D}(D-g(v))^2f_{V}(v)dv.
\end{align}

\subsubsection{$\mathbb{E}[Q_{i}|B]$} 
Since only the first packet that arrives during the busy period is serviced and others are discarded, we introduce a lemma for the probability that an arriving packet is the first one that arrives in the busy state. To do so, we first define the states $B_{1}$ and $B_2$ as the busy states of the server with zero and one packet waiting in the queue, respectively. Recall the renewal cycle: After idle period, an arrival happens and system turns to $B_{1}$ state. Now a time duration of service $S$ starts and if during the service period another arrival occurs, the system turns to $B_{2}$ state. This back-and-forth between $B_1$ and $B_2$ states continues until no packet arrives in one service time. This renewal structure yields the following result:

\begin{Lemma}
	 In M/GI/1/2 scheme, waiting time of a packet in the buffer conditioned on its arrival in $B_1$ state is:
	\begin{align*}
	\mathbb{E}[W_{B_{2}}]&=\mathbb{E}[S-X|X<S]Pr[X<S]\\
	&=\mathbb{E}[S]+\frac{1}{\lambda}MGF_{S}(\lambda)-\frac{1}{\lambda}.
	\end{align*}
	The stationary probability of $B_{2}$ state is:
	\begin{align*}
	p_{B_{2}}=p_{B}\frac{\mathbb{E}[W_{B_{2}}]}{\mathbb{E}[S]}=p_{B}(1+\frac{MGF_{S}(\lambda)-1}{\lambda \mathbb{E}[S]}),
	\end{align*}
	and the probability of $B_{1}$ state is $p_{B_{1}}=p_{B}-p_{B_{2}}$.
\end{Lemma}	 
Then we have $\mathbb{E}[Q_{i}|B]=\mathbb{E}[Q_{i}|B_{1}]$ and we give the probability distribution function for the conditional residual service time $W^{'}$under the condition that the packet arrives in $B_{1}$ state: 
\begin{align}
\nonumber \mathbb{P}[W^{'}>w]=\mathbb{P}[S-X>w|X<S]\\
\nonumber =\frac{\int_{w}^{\infty}\int_{0}^{s-w}f_{S}(s)f_{X}(x)dxds}{\mathbb{P}[X<S]}\\
 =\frac{\int_{w}^{\infty}f_{S}(s)(1-e^{-\lambda(s-w)})ds}{1-MGF_{S}(\lambda)}, \nonumber
\end{align}
and we have:
\begin{align}
f_{W^{'}}(w)=\frac{d(1-\mathbb{P}[W^{'}>w])}{dw}.
\end{align}
\begin{align}
\nonumber Q_{i}=\left\{\begin{matrix}
\frac{V_{0,i}}{2D}(D-(S_{i}+W^{'}_{i-1}))^2 &(S_{i}+W^{'}_{i-1}<D) \\ 
0 \ \ &(S_{i}+W^{'}_{i-1}>D) 
\end{matrix}\right.\\
\nonumber \mathbb{E}[Q_{i}|B]=\int_{0}^{\tilde{V}}\int_{0}^{D-g(v)}Q_{i}f_{W^{'}}(w)f_{V}(v)dwdv.
\end{align}
Therefore, we have $\mathbb{E}[Q_{i}]=\mathbb{E}[Q_{i}|I]p_{I}+\mathbb{E}[Q_{i}|B_{1}]p_{B_{1}}$ and average value is $\mathbb{E}[VoI]=\lambda\mathbb{E}[Q_{i}]$.

\subsection{Average Value for $M/GI/1/2^*$}

The $M/GI/1/2^*$ queueing system is the same as M/GI/1/2 except that we use last come first serve with packet discarding. The latest packet arriving in a busy period takes the place of the old packet in the buffer. Therefore, we have the same stationary probabilities for each state as the $M/GI/1/2$ system in (\ref{prob:MG12}). Additionally, the expression for $\mathbb{E}[Q_{i}|I]$ is the same as (\ref{EQ,I:MG12}). Next we evaluate the expression for $\mathbb{E}[Q_{i}|B]$.
\subsubsection{$\mathbb{E}[Q_{i}|B]$}
If the $i$th packet arrives to the server during the busy period, it will be transmitted to the destination conditioned on the event $\{X_{i}>W_{i-1}\}$ which means the next packet arrives to the server after the current service finishes. Then we have:
\begin{align}
\nonumber Q_{i}=\left\{\begin{matrix}
\frac{V_{0,i}}{2D}(D-(S_{i}+W_{i-1}))^2 &(S_{i}+W_{i-1}<D) \\ 
0 \ \ &(S_{i}+W_{i-1}>D). 
\end{matrix}\right.
\end{align}

\begin{align}
\nonumber \mathbb{E}[Q_{i}|B]=\int_{0}^{\tilde{V}}\int_{0}^{D-g(v)}\int_{w}^{\infty}Q_{i}f_{X}(x)f_{W}(w)f_{V}(v)dxdwdv.
\end{align}
Therefore, we have $\mathbb{E}[Q_{i}]=\mathbb{E}[Q_{i}|I]p_{I}+\mathbb{E}[Q_{i}|B]p_{B}$, and $\mathbb{E}[VoI]=\lambda\mathbb{E}[Q_{i}]$.

\section{Numerical Results}

In this section, we provide numerical results for average VoI for various cases. We also perform packet-based queue simulations offline for $10^6$ packets as verification of the results. The lifetime of packets is fixed to $D=3$ throughout.

\subsection{Uniformly Distributed Initial Value}

First, we assume the initial value of each packet is uniformly distributed between $V_{min}$ and $V_{max}$. We use $g(V)=a \log(V+1)$ as relation between value and service time representing an exponential valuation of time, typical of distortion-based quality of information scenarios \cite{rajaraman2018not,bastopcu2019,ayan2019age}. In Appendix A, we provide closed form expressions for this case. 

\begin{figure}[!t]
	\centering{
		\hspace{-0.2cm} 
		\includegraphics[totalheight=0.26\textheight]{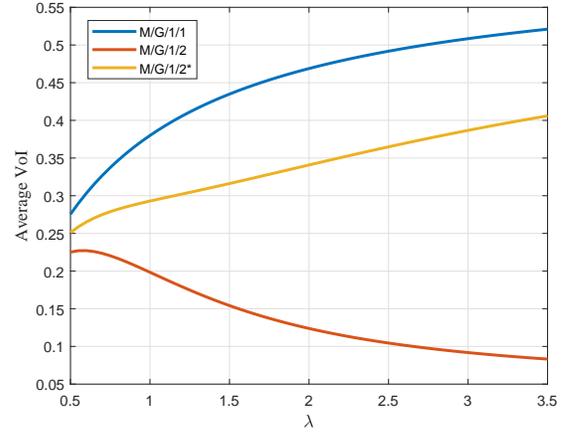}}\vspace{-0.15in}
	\caption{\sl Average VoI for uniform distribution with respect to $\lambda$ for $V_{min}=0$, $V_{max}=10$, $a=1$.}
	\label{VoI:Uniform} 
	\vspace{-0.2in}
\end{figure}

In Fig. \ref{VoI:Uniform}, we show average VoI versus arrival rate $\lambda$ for three queuing schemes. We observe that M/GI/1/1 performs better than the other two and this is in sharp contrast with a similar comparison in \cite{costa2016age}. In particular, due to the exponential relation between time and value, keeping a packet in the buffer to keep the server busy turns out to yield smaller value at the receiver with respect to keeping none and serving only the freshest packets. For M/GI/1/2, on the other hand, there is an optimal point with respect to $\lambda$ after which average VoI drops due to undesired increases in waiting times in the data buffer.

%

%
%

\subsection{Exponentially Distributed Initial Value}

Next, we consider $f_V(v)=\mu e^{-\mu v}$ and the service time is $g(V)=V$. This is an initial value representing the number of bits served to the receiver with a linear descend in time. In Appendix B, we provide closed form expressions for $\mathbb{E}[VoI]$.

\begin{figure}[!t]
	\centering{
		\hspace{-0.2cm} 
		\includegraphics[totalheight=0.26\textheight]{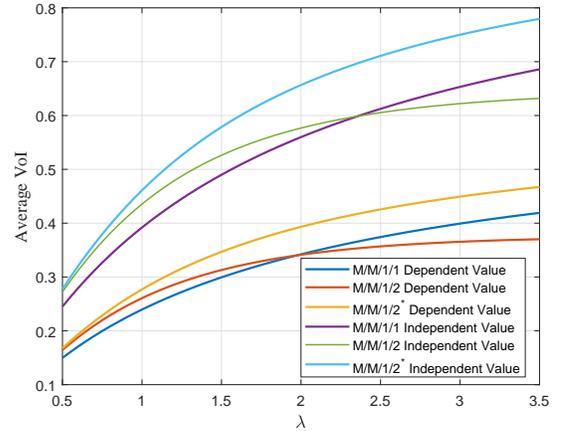}}\vspace{-0.15in}
	\caption{\sl Average VoI for exponentially distributed service time and value.}
	\label{VoI:independt M/M/1} \vspace{-0.2in}
\end{figure}

In Fig. \ref{VoI:independt M/M/1}, we set $\mu=1.5$ and plot average VoI versus $\lambda$. We also show average VoI for independent initial value and service time under same marginal distributions. We observe that independent service time yields higher value as the adverse relation between initial value and service rate is eradicated.

\subsection{Binary Distributed Initial Value}

We finally consider binary distributed initial value for two classes of update packets. Class 1 and class 2 packets have $V_{0,i}=V_{1}$ and $V_{0,i}=V_{2}$. Whether $i$th packet is in class 1 or 2 is independent over $i$ and with probability $p$, $(1-p)$ respectively. This selection models a case when a packet contains message about an alarming event yielding high value once received and other type of packets are assumed regular. 

\begin{figure}[!t]
 	\centering{
 		\hspace{-0.2cm} 
 		\includegraphics[totalheight=0.26\textheight]{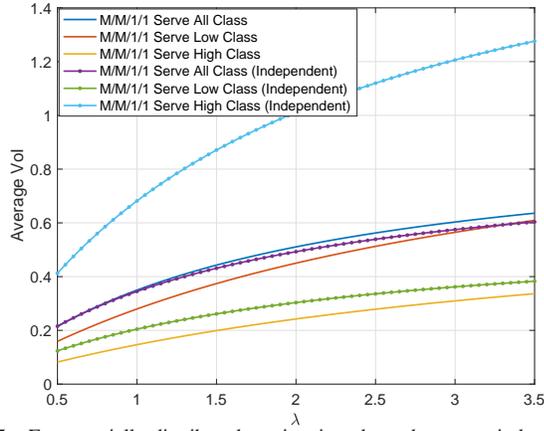}}\vspace{-0.2in}
 	\caption{\sl Exponentially distributed service time dependent on or independent of the binary value in M/M/1/1 scheme.}\vspace{-0.15in}
 	\label{VoI:dependt Binary} 
 \end{figure}

In Fig. \ref{VoI:dependt Binary}, we set $V_{1}=0.4$, $V_{2}=1.33$ and $p=0.8$, $\mu=1.5$. We compare plots showing average VoI versus $\lambda$ for three different policies. The first policy serves all packets without regard to the value, the second policy involves serving only class 1 packets and the third policy serves only class 2 packets. Our numerical results show that when service time is independent of value, always serving the high-value packet will yield the highest average value. On the other hand, in the dependent case when arrival rate becomes large, serving the packet with low value but smaller service time and high probability will benefit the average VoI compared to serving all the packets or serving the high-value packets with larger service time and low probability.

\begin{figure}[t] 
	\centering \vspace{-0.0in}
	\subfigure[]
	{
		\includegraphics[totalheight=0.26\textheight]{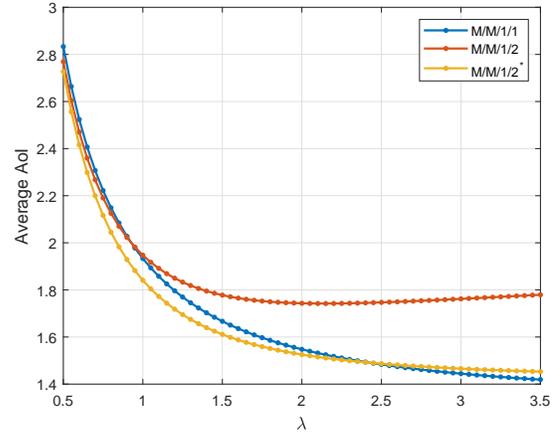}\vspace{-0.22in} 
		\label{fig:local1}}\vspace{-0.07in} 
	\subfigure[]
	{
		\includegraphics[totalheight=0.26\textheight]{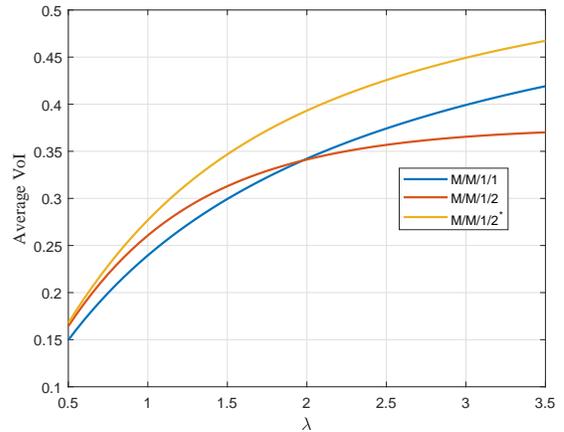} \vspace{-0.15in}
		\label{fig:local2}}\vspace{-0.1in}
	
	\caption{\sl Average AoI and average VoI versus $\lambda$ for M/M/1/1, M/M/1/2 and M/M/1/$2^*$ schemes. (a) Average AoI, (b) Average VoI.  }\vspace{-0.25in}
	\label{fig:local001} 
\end{figure}

We close numerical results with a comparison of average VoI and average AoI as shown in Fig. \ref{fig:local001} where plots are drawn with respect to arrival rate $\lambda$. In these plots, the initial value is exponentially distributed, $g(V)=V$ and $\mu=1.5$. In Fig. \ref{fig:local1}, we use the expressions from \cite{costa2016age} for the average AoI in various systems. We can see in this figure that when the arrival rate is small, the AoIs of all schemes are similar. As the arrival rate increases, M/M/1/$2^{*}$ and M/M/1/1 schemes are better than the other with a crossing as $\lambda$ tends to infinity. When $\lambda$ is large, M/M/1/1 performs better than the M/M/1/$2^{*}$ system. Note also that for M/M/1/2, average AoI is not monotonic and there is an optimal $\lambda$ for this scheme that optimizes the average AoI.

In Fig. \ref{fig:local2}, we plot the average VoI with respect to $\lambda$ for various schemes. In contrast to the comparisons for average AoI, we observe that M/M/1/$2^*$ always performs better than the others. This is connected to the fact that VoI is the accumulated value of received packet values so that the total value is higher if a packet is stored in the buffer instead of dropping it. As the arrival rate $\lambda$ increases, M/M/1/2 performs the worst in terms of average VoI as in the case of AoI. 

\section{Conclusion}

Age of information (AoI) is a well-known metric that quantifies the freshness of information at a receiver in status update systems. This metric ignores the potential differences in the importance of the various update packets. In this paper, we consider the value of information in status update systems wherein packets have various initial values upon generation. We investigate various queuing disciplines with initial-value-dependent packet service times, and obtain closed-form expressions for average VoI. Our numerical results illustrate the average VoI for different scenarios and the contrast between AoI and VoI. In particular, AoI promotes latest updates to be received by the receiver while VoI measures an additive reward over packets. We observe the role of dependence between service time and initial value in the average VoI. We additionally compare differing effects of data buffer and increased packet arrival rate on average VoI and average AoI. 


\appendix

\subsection {$\mathbb{E}[VoI]$ for M/G/1/1 with Uniformly Distributed Value}  
In uniform case, we have $f_{V}(v)=\frac{1}{u}$ and we assume $g(V)=a \mbox{log}( 1 + V)$. So we have the mean service time:
\begin{align*}
&\mathbb{E}[S]=\mathbb{E}[a \log(1+V)]=\int_{V_{min}}^{V_{max}}a \log(1+v)f_V(v)dv\\
&=\frac{a}{u}\int_{V_{min}}^{V_{max}} \log(1+v)dv \end{align*} \begin{align*} &=\frac{a}{u}((V_{max}+1)\log(V_{max}+1)-(V_{min}+1)\log(V_{min}+1)).
\end{align*} 
So from (\ref{MG11:pb}), we have
\begin{align*}
&p_I=\\&\frac{1}{1+\lambda\frac{a}{u}((V_{max}+1)\log(V_{max}+1)-(V_{min}+1)\log(V_{min}+1))}.
\end{align*} 
Then we calculate $\mathbb{E}[Q_i]$ from (\ref{EQi:MG11}) and we have:
\begin{align*}
\mathbb{E}[Q_{i}]=\frac{p_{I}}{2}\int_{0}^{V^{'}_{i}}\frac{v}{D}(D-g(v))^2f_{V}(v)dv.
\end{align*}
Here, $a \log(\tilde{V}+1)=D$ so we have $\tilde{V}=e^{\frac{D}{a}}-1$. Define $V_{up}=\tilde{V}$ if $\tilde{V}<V_{max}$ and $V_{up}=V_{max}$ otherwise. Then:
\begin{align*}
\mathbb{E}[Q_{i}]=&\frac{p_{I}}{2}\int_{V_{min}}^{V_{up}}\frac{v}{D}(D-a\log(v))^2f_{V}(v)dv\\
=&\frac{p_{I}}{2Du}\int_{V_{min}}^{V_{up}}D^2v-2aDv\log(v)+v(a\log(v))^2dv\\
=&\frac{p_{I}}{2Du}\bigg(D^2\frac{V_{up}^2-V^2_{min}}{2} \end{align*}\begin{align*}
&-\frac{aD}{2}(V_{up}^2(2\log(V_{up})-1)-V_{min}^2(2\log(V_{min})-1))\\
&+\frac{a^2}{4}(V_{up}^2(2\log(V_{up})^2-2\log(V_{up})+1))\\
&-(V_{min}^2(2\log(V_{min})^2-2\log(V_{min})+1))\bigg).
\end{align*}
Finally we have $\mathbb{E}[VoI]=\lambda \mathbb{E}[Q_{i}]$.
\subsection {$\mathbb{E}[VoI]$ for M/M/1/2 with Exponentially Distributed Value}  
\hspace{-0.2in}For $f_V(v)=\mu e^{-\mu v}$ $g(V)=V$, we have $\tilde{V}=D$ and from (\ref{EQ,I:MG12}):
\begin{align*}
\mathbb{E}[Q_{i}|I]&=\frac{1}{2}\int_{0}^{D}\frac{v}{D}(D-v)^2f_{V}(v)dv\\
&=\frac{1}{2D}\int_{0}^{D}(D^2v-2Dv^2+v^3)f_{V}(v)dv.
\end{align*}
Evaluating this integral, we obtain the expression:
\begin{align*}
\mathbb{E}[Q_{i}|I]=\frac{D^2\mu^2-4D\mu +6-e^{-D\mu}(2D\mu+6)}{2D\mu^3}.
\end{align*}
Also, since $W$ is exponentially distributed with $\mu$, we get:
\begin{align*}
\mathbb{E}[Q_{i}|B]=&\int_{0}^{D}\int_{0}^{D-v}\frac{v}{2D}(D-(v+w))^2f_{W}(w)f_{V}(v)dwdv\\
=&-\frac{e^{-D\mu}(6D\mu+D^2\mu^2+12)}{2D\mu^3}\\
&+\frac{12+D^2\mu^2-6D\mu}{2D\mu^3}.
\end{align*}
From (\ref{prob:MG12}) and Lemma 1, we have $p_I=\frac{\mu^2}{\lambda^2+\lambda\mu+\mu^2}$ and $p_{B1}=\frac{\lambda\mu}{\lambda^2+\lambda\mu+\mu^2}$. We get $\mathbb{E}[VoI]=\lambda(\mathbb{E}[Q_{i}|I]p_{I}+\mathbb{E}[Q_{i}|B]p_{B1})$. 

\vspace{0.1in}

\begin{thebibliography}{10}
	
	\bibitem{giordani2019investigating}
	M.~Giordani, A.~Zanella, T.~Higuchi, O.~Altintas, and M.~Zorzi.
	\newblock Investigating value of information in future vehicular
	communications.
	\newblock {\em arXiv preprint arXiv:1907.10124}, 2019.
	
	\bibitem{higuchi2019value}
	T.~Higuchi, M.~Giordani, A.~Zanella, M.~Zorzi, and O.~Altintas.
	\newblock Value-anticipating v2v communications for cooperative perception.
	\newblock In {\em 2019 IEEE Intelligent Vehicles Symposium}, pages 1947--1952,
	2019.
	
	\bibitem{giordani2019framework}
	M.~Giordani, T.~Higuchi, A.~Zanella, O.~Altintas, and M.~Zorzi.
	\newblock A framework to assess value of information in future vehicular
	networks.
	\newblock In {\em ACM MobiHoc Workshop on Technologies, mOdels, and Protocols
		for Cooperative Connected Cars}, pages 31--36, 2019.
	
	\bibitem{boloni2013scheduling}
	L.~B{\"o}l{\"o}ni, D.~Turgut, S.~Basagni, and C.~Petrioli.
	\newblock Scheduling data transmissions of underwater sensor nodes for
	maximizing value of information.
	\newblock In {\em IEEE GLOBECOM}, pages 438--443, 2013.
	
	\bibitem{turgut2013ive}
	D.~Turgut and L.~B{\"o}l{\"o}ni.
	\newblock Ive: Improving the value of information in energy-constrained
	intruder tracking sensor networks.
	\newblock In {\em 2013 IEEE ICC}, pages 6360--6364, 2013.
	
	\bibitem{bidoki2018joint}
	N.H. Bidoki, M.B. Baghdadabad, G.~Sukthankar, and D.~Turgut.
	\newblock Joint value of information and energy aware sleep scheduling in
	wireless sensor networks: A linear programming approach.
	\newblock In {\em IEEE ICC}, 2018.
	
	\bibitem{suri2015exploring}
	N.~Suri, G.~Benincasa, R.~Lenzi, M.~Tortonesi, C.~Stefanelli, and L.~Sadler.
	\newblock Exploring value-of-information-based approaches to support effective
	communications in tactical networks.
	\newblock {\em IEEE Communications Magazine}, 53(10):39--45, 2015.
	
	\bibitem{kamar2013light}
	E.~Kamar and E.~Horvitz.
	\newblock Light at the end of the tunnel: A monte carlo approach to computing
	value of information.
	\newblock In {\em International conference on Autonomous agents and multi-agent
		systems}, pages 571--578, 2013.
	
	\bibitem{rosenthal2013look}
	S.~Rosenthal, D.~Bohus, E.~Kamar, and E.~Horvitz.
	\newblock Look versus leap: computing value of information with
	high-dimensional streaming evidence.
	\newblock In {\em Twenty-Third International Joint Conference on Artificial
		Intelligence}, 2013.
	
	\bibitem{kaul2012real}
	S.~Kaul, R.~Yates, and M.~Gruteser.
	\newblock Real-time status: How often should one update ?
	\newblock In {\em INFOCOM}, pages 2731--2735. IEEE, 2012.
	
	\bibitem{kaul2011minimizing}
	S.~Kaul, M.~Gruteser, V.~Rai, and J.~Kenney.
	\newblock Minimizing age of information in vehicular networks.
	\newblock In {\em IEEE Conference on Sensor, Mesh and Ad Hoc Communications and
		Networks}, pages 350--358, 2011.
	
	\bibitem{costa2016age}
	M.~Costa, M.~Codreanu, and A.~Ephremides.
	\newblock On the age of information in status update systems with packet
	management.
	\newblock {\em IEEE Transactions on Information Theory}, 62(4):1897--1910,
	2016.
	
	\bibitem{najm2016age}
	E.~Najm and R.~Nasser.
	\newblock Age of information: The gamma awakening.
	\newblock In {\em Information Theory (ISIT), 2016 IEEE International Symposium
		on}, pages 2574--2578. Ieee, 2016.
	
	\bibitem{inoue2018general}
	Y.~Inoue, H.~Masuyama, T.~Takine, and T.~Tanaka.
	\newblock A general formula for the stationary distribution of the age of
	information and its application to single-server queues.
	\newblock {\em arXiv preprint arXiv:1804.06139}, 2018.
	
	\bibitem{2018information}
	A.~Baknina, O.~Ozel, J.~Yang, S.~Ulukus, and A.~Yener.
	\newblock Sending information through status updates.
	\newblock In {\em IEEE ISIT}, 2018.
	
	\bibitem{bacinoglu2015age}
	B.~T. Bacinoglu, E.~T. Ceran, and E.~Uysal-Biyikoglu.
	\newblock Age of information under energy replenishment constraints.
	\newblock In {\em USCD ITA}, February 2015.
	
	\bibitem{yates2015lazy}
	R.~Yates.
	\newblock Lazy is timely: Status updates by an energy harvesting source.
	\newblock In {\em IEEE ISIT}, June 2015.
	
	\bibitem{wu2017optimal_ieee}
	X.~Wu, J.~Yang, and J.~Wu.
	\newblock Optimal status update for age of information minimization with an
	energy harvesting source.
	\newblock {\em IEEE Trans. on Green Communications and Networking}, 2(1), March
	2018.
	
	\bibitem{bacinoglu2017scheduling}
	B.T. Bacinoglu and E.~Uysal-Biyikoglu.
	\newblock Scheduling status updates to minimize age of information with an
	energy harvesting sensor.
	\newblock In {\em IEEE ISIT}, pages 1122--1126, 2017.
	
	\bibitem{farazi2018average}
	S.~Farazi, A.G. Klein, and D.R. Brown.
	\newblock Average age of information for status update systems with an energy
	harvesting server.
	\newblock In {\em IEEE INFOCOM WORKSHPS}, pages 112--117, 2018.
	
	\bibitem{arafa2017age}
	A.~Arafa and S.~Ulukus.
	\newblock Age-minimal transmission in energy harvesting two-hop networks.
	\newblock In {\em IEEE Globecom}, December 2017.
	
	\bibitem{talak2017minimizing}
	R.~Talak, S.~Karaman, and E.~Modiano.
	\newblock Minimizing age-of-information in multi-hop wireless networks.
	\newblock In {\em Communication, Control, and Computing (Allerton), 2017 55th
		Annual Allerton Conference on}, pages 486--493. IEEE, 2017.
	
	\bibitem{kadota2018optimizing}
	I.~Kadota, A.~Sinha, and E.~Modiano.
	\newblock Optimizing age of information in wireless networks with throughput
	constraints.
	\newblock In {\em IEEE INFOCOM}, pages 1844--1852, 2018.
	
	\bibitem{yates2018age}
	R.D. Yates.
	\newblock The age of information in networks: Moments, distributions, and
	sampling.
	\newblock {\em arXiv preprint arXiv:1806.03487}, 2018.
	
	\bibitem{maatouk2018age}
	A.~Maatouk, M.~Assaad, and A.~Ephremides.
	\newblock The age of updates in a simple relay network.
	\newblock {\em arXiv preprint arXiv:1805.11720}, 2018.
	
	\bibitem{Alabbasi2018JointIF}
	A.~Alabbasi and V.~Aggarwal.
	\newblock Joint information freshness and completion time optimization for
	vehicular networks.
	\newblock {\em CoRR}, abs/1811.12924, 2018.
	
	\bibitem{rajaraman2018not}
	N.~Rajaraman, R.~Vaze, and G.~Reddy.
	\newblock Not just age but age and quality of information.
	\newblock {\em arXiv preprint arXiv:1812.08617}, 2018.
	
	\bibitem{AKosta2017}
	A.~{Kosta}, N.~{Pappas}, A.~{Ephremides}, and V.~{Angelakis}.
	\newblock Age and value of information: Non-linear age case.
	\newblock In {\em 2017 IEEE International Symposium on Information Theory
		(ISIT)}, pages 326--330, June 2017.
	
	\bibitem{maatouk2019age}
	A.~Maatouk, S.~Kriouile, M.~Assaad, and A.~Ephremides.
	\newblock The age of incorrect information: A new performance metric for status
	updates.
	\newblock {\em arXiv preprint arXiv:1907.06604}, 2019.
	
	\bibitem{bastopcu2019}
	M.~Bastopcu and S.~Ulukus.
	\newblock Age of information for updates with distortion.
	\newblock In {\em IEEE ITW}, 2019.
	
	\bibitem{ayan2019age}
	O.~Ayan, M.~Vilgelm, M.~Kl{\"u}gel, S.~Hirche, and W.~Kellerer.
	\newblock Age-of-information vs. value-of-information scheduling for cellular
	networked control systems.
	\newblock In {\em ACM/IEEE International Conference on Cyber-Physical Systems},
	pages 109--117, 2019.
	
	\bibitem{CKam2016MILCOM}
	C.~{Kam}, S.~{Kompella}, G.~D. {Nguyen}, J.~E. {Wieselthier}, and
	A.~{Ephremides}.
	\newblock Controlling the age of information: Buffer size, deadline, and packet
	replacement.
	\newblock In {\em IEEE MILCOM}, pages 301--306, 2016.
	
	\bibitem{CKam2016ISIT}
	C.~{Kam}, S.~{Kompella}, G.~D. {Nguyen}, J.~E. {Wieselthier}, and
	A.~{Ephremides}.
	\newblock Age of information with a packet deadline.
	\newblock In {\em IEEE ISIT}, pages 2564--2568, 2016.
	
	\bibitem{CKam2018}
	C.~{Kam}, S.~{Kompella}, G.~D. {Nguyen}, J.~E. {Wieselthier}, and
	A.~{Ephremides}.
	\newblock On the age of information with packet deadlines.
	\newblock {\em IEEE Transactions on Information Theory}, 64(9):6419--6428, Sep.
	2018.
	
	\bibitem{YInoue2018}
	Y.~{Inoue}.
	\newblock Analysis of the age of information with packet deadline and infinite
	buffer capacity.
	\newblock In {\em IEEE ISIT}, pages 2639--2643, 2018.

\end{thebibliography}
\end{document}